\documentstyle[11pt,epsfig]{article}

\begin{document}

\begin{center}
{\large {\bf ''HOMOGENEOUS'' GRAVITATIONAL FIELD \\[2mm]
IN GENERAL RELATIVITY ?}}

\bigskip

{\bf R. M. Avakyan, E. V. Chubaryan, A. H. Yeranyan}

\bigskip

{\it Department of Physics, Yerevan State University}

{\it 1 Alex Manoogian St., 375049 Yerevan, Armenia \\
E-mail ayeran@www.physdep.r.am }
\end{center}

\bigskip

{\small The gravitation field of the flat plate was investigated. It have
been shown that there exist the internal solution of Einstein equations
sewed together with external one, which described a ''homogeneous''
gravitational field.}

\bigskip

{\large Introduction.} The interest to the problems with planar symmetry has
again increased in the last years. First of all it is connected to the fact
that this type of solution plays an important role in the theories of
superstrings and supergravity \cite{SDW}.

There are series of papers devoted to the study of gravitational field of a
flat plate (see Ref. in \cite{gron}). The external solution of this problem
have been found for the first time by Taub \cite{TB}. The numerical
integration of internal solution and its sewing together with external one
was performed in \cite{RMG}. Here, we stress that Einstein equations accept
two types of internal solutions: with negative and positive derivative of
metric tensor at the centre of plate.

We will show in this paper, that only solution with negative derivative can
be sewing together Taub solution. {\it The second one corresponds to
gravitational field for which the Riamannian tensor is zero outside of the
plate.}

{\large Flat plate: general.} Let's consider a static gravitational field of
infinite flat plate. Following the paper ~\cite{RMG}, space - time metric
can be written as (axis $Ox$ is perpendicular to the plate, and the plane $%
yz $ is located at the middle of the plate)
\begin{equation}
ds^2=e^{\nu \left( x\right) }dt^2-dx^2-e^{\lambda \left( x\right) }\left(
dy^2+dz^2\right) .  \label{metric}
\end{equation}

Obviously, $\nu \left( x\right) $ and $\lambda \left( x\right) $ are even
functions of $x$. Below for definiteness we shall consider the solutions in $%
x>0$.\

Einstein's equations and equations of hydrodynamics (with energy-momentum
tensor for ideal liquid) appropriate to the metric (\ref{metric}) are as
follows (using system of units $c=G=1$):

\begin{equation}
\lambda ^{\prime \prime }+\frac 34\lambda ^{\prime 2}=-8\pi \rho ,
\label{eq1}
\end{equation}

\begin{equation}
\frac{\lambda ^{\prime 2}}4+\frac{\lambda ^{\prime }\nu ^{\prime }}2=8\pi p,
\label{eq2}
\end{equation}

\begin{equation}
\frac{\nu ^{\prime \prime }}2+\frac{\nu ^{\prime 2}}4+\frac{\lambda ^{\prime
\prime }}2+\frac{\lambda ^{\prime 2}}4+\frac{\lambda ^{\prime }\nu ^{\prime }%
}4=8\pi p,  \label{eq3}
\end{equation}

\begin{equation}
p^{\prime }=-\frac{\nu ^{\prime }}2\left( p+\rho \right) ,  \label{eq4}
\end{equation}
where $\rho (x)$ is the energy density, $p(x)$ is the pressure, and prime
denotes the differentiation with respect to $x$. We should add to this
system the state equation $p=p(\rho )$. We would like to remind that one of
the equations of system (\ref{eq1}) - (\ref{eq4}) is a consequence of the
others.

As boundary conditions for system (\ref{eq1}) - (\ref{eq4}) we have to
specify values for $\nu (0)$, $\nu ^{\prime }(0)$, $\lambda (0)$, $\lambda
^{\prime }(0)$ and $p(0)$ at the centre. The functions $\nu \left( x\right) $
and $\lambda \left( x\right) $ are determined to up additive constants (as
they are not contained in the equations evidently), which can be removed by
simple scale transformation: $t\rightarrow \alpha t$, $y\rightarrow \beta y$%
, $z\rightarrow \beta z$ . Therefore, we can put

\begin{equation}
\nu (0)=\lambda (0)=0.  \label{boundcond0}
\end{equation}
It is obvious from symmetry, that derivative of pressure (force) at the
centre of the configuration is equal to zero, i.e.

\begin{equation}
p^{\prime }(0)=0,  \label{boundcond1}
\end{equation}
therefore

\begin{equation}
\nu ^{\prime }(0)=0.  \label{boundcond2}
\end{equation}
Inside the configuration pressure decreases monotonically ($p^{\prime }<0$
), and $\nu \left( x\right) $ increases monotonically ($\nu ^{\prime }>0$ ).
The bound of configuration $x_s$ is determined by condition

\begin{equation}
p(x_s)=0.  \label{bound}
\end{equation}
The derivative of $\lambda $ at the centre of the plate is determined from
the equation (\ref{eq2})

\begin{equation}
\lambda ^{\prime }(0)=\pm \sqrt{32\pi p_0},  \label{boundcond3}
\end{equation}
where $p_0\equiv p(0)$ . At $x=x_s$ the pressure is equal to zero,
therefore, as it follows from the equation (\ref{eq2}), $\lambda ^{\prime
}(x_s)=0$ or $\lambda ^{\prime }(x_s)=-2\nu ^{\prime }(x_s)<0$. Inside the
configuration the function $e^{\lambda (x)}$ is monotone and should not have
singularity, therefore if $\lambda ^{\prime }(0)=\sqrt{32\pi p_0}$ then $%
\lambda ^{\prime }(x_s)=0$ and $\lambda ^{\prime \prime }(x_s)=0$ (from the
equation (\ref{eq1})), if $\lambda ^{\prime }(0)=-\sqrt{32\pi p_0}$ then $%
\lambda ^{\prime }(x_s)=-2\nu ^{\prime }(x_s)$. According to this, depending
on sign of $\lambda ^{\prime }(0)$ the internal solution will be sewed
either with the external solution of Taub \cite{TB} (for $\lambda ^{\prime
}(0)=-\sqrt{32\pi p_0}$)

\begin{equation}
ds^2=\frac A{\left( 1-Bx\right) ^{\frac 23}}dt^2-dx^2-C\left( 1-Bx\right)
^{\frac 43}\left( dy^2+dz^2\right) ,  \label{Taybsol}
\end{equation}
where $A>0$, $B$ and $C>0$ are constants of integration, or with external
solution distinct essentially from it (for $\lambda ^{\prime }(0)=\sqrt{%
32\pi p_0}$)

\begin{equation}
ds^2=\left( a+bx\right) ^2dt^2-dx^2-c\left( dy^2+dz^2\right) ,
\label{oursol}
\end{equation}
where $a>0$, $b$ and $c>0$ are also constants of integration.

{\large Model of ideal liquid.} We study in detail the case of homogeneous
liquid $\rho =\rho _0=const$. The equation (\ref{eq1}) can be integrated
easily and taking into account the boundary conditions (\ref{boundcond0})
and (\ref{boundcond3}) we have:

\begin{equation}
e^{\lambda \left( x\right) }=\left( 1+3q_0\right) ^{\frac 23}\left[ \cos
\left( \sqrt{6\pi \rho _0}x\pm \arccos \frac 1{\sqrt{1+3q_0}}\right) \right]
^{\frac 43},  \label{lambda}
\end{equation}
where $q_0\equiv p_0/\rho _0$ the ratio of the central pressure to the
density. In the formula (\ref{lambda}) upper sign corresponds to the
solution with boundary condition $\lambda ^{\prime }(0)=-\sqrt{32\pi p_0}$,
and lower sign corresponds to $\lambda ^{\prime }(0)=\sqrt{32\pi p_0}$. As
in the second case $\lambda ^{\prime }(x_s)=0$, from (\ref{lambda}) we can
find also the bound of the configuration $x_s$. Differentiating the right
hand side of (\ref{lambda}) and equating to zero we find:

\begin{equation}
x_s=\frac{\arccos \frac 1{\sqrt{1+3q_0}}}{\sqrt{6\pi \rho _0}}.  \label{boun}
\end{equation}
In the second case the value of $\lambda $ on bound is equal to

\begin{equation}
\lambda (x_s)=\frac 23\ln \left( 1+3q_0\right) .  \label{lambdas}
\end{equation}

The equation of hydrodynamics can be also integrated

\begin{equation}
p=\left( p+\rho \right) e^{-\frac \nu 2}-\rho _0.  \label{pres}
\end{equation}
Therefore, in both cases the value of $\nu $ on bound is equal to

\begin{equation}
\nu \left( x_s\right) =2\ln \left( 1+q_0\right) .  \label{nwus}
\end{equation}

Substituting (\ref{lambda}) and (\ref{pres}) into (\ref{eq2}), and taking
into account the boundary conditions we find:

\begin{eqnarray}
e^{\nu (x)} & = & ( 1 + q_0) ^2\left[ 1 + \frac{\sin \left( \sqrt{6\pi \rho
_0}x\pm \arccos \frac 1{\sqrt{1+3q_0}}\right) }{\cos ^{\frac 13}\left( \sqrt{%
6\pi \rho _0}x\pm \arccos \frac 1{\sqrt{1+3q_0}}\right) }\left( \mp \sqrt{%
\frac{q_0}3}\frac{\left( 1+3q_0\right) ^{\frac 13}}{\left( 1+q_0\right) }%
+\right. \right.  \nonumber \\
&& \left. \left. +\frac{\sqrt{6\pi \rho _0}}3 \int_0^x\frac{dx}{\cos ^{\frac
23}\left( \sqrt{6\pi \rho _0}x\pm \arccos \frac 1{\sqrt{1+3q_0}}\right) }%
\right) \right]  \label{nwu}
\end{eqnarray}

To find the constants of integration $A$, $B$ and $C$ or $a$, $b$ and $c$ we
should require continuity of the components of metric tensor and their first
derivative on the bound of the configuration. However, as (\ref{pres}) (with
the account (\ref{nwu})) is rather complicated, the bound of the
configuration could not be found explicitly, and sewing together can not be
made analytically. Therefore it's convenient to find the internal solution
by numerical calculation. For this purpose we should write down system of
the equations (\ref{eq1}) - (\ref{eq4}) in the convenient form. It's
reasonable, for that, besides $\nu \left( x\right) $, $\lambda \left(
x\right) $ and $p\left( x\right) $ to calculate the following functions

\begin{equation}
\sigma \left( x\right) =2\int_0^x\left( T_0^0-T_1^1-T_2^2-T_3^3\right) \sqrt{%
-g}dx=2\int_0^x\left( \rho +3p\right) e^{\frac \nu 2+\lambda }dx,
\label{masa}
\end{equation}

\begin{equation}
\chi \left( x\right) =2\int_0^x\left( T_0^0+T_1^1\right) \sqrt{-g}%
dx=2\int_0^x\left( \rho -p\right) e^{\frac \nu 2+\lambda }dx.  \label{dmasa}
\end{equation}
The function $\sigma \left( x\right) $ is ''accumulated'' surface density
(it can be easily verified passing to the Newtonian limit [1]). Taking into
account (\ref{masa}), (\ref{dmasa}) and replacing the independent variable

\begin{equation}
x=\frac 1{\sqrt{8\pi \rho _0}}\stackrel{\sim }{x},  \label{coord}
\end{equation}
we can write equations (\ref{eq1}) - (\ref{eq4}) in the following form:

\begin{equation}
\stackrel{\stackrel{\cdot }{\sim }}{\sigma }=2\left( 1+3q\right) e^{\frac
\nu 2+\lambda },  \label{equt1}
\end{equation}

\begin{equation}
\stackrel{.}{\nu }=\frac 12\stackrel{\sim }{\sigma }e^{-\frac \nu 2-\lambda
},  \label{equt2}
\end{equation}

\begin{equation}
\stackrel{\stackrel{\cdot }{\sim }}{\chi }=2\left( 1-q\right) e^{\frac \nu
2+\lambda },  \label{equt3}
\end{equation}

\begin{equation}
\stackrel{.}{\lambda }=-\frac 12\stackrel{\sim }{\chi }e^{-\frac \nu
2-\lambda },  \label{equt4}
\end{equation}

\begin{equation}
\stackrel{.}{q}=-\frac 14\stackrel{\sim }{\sigma }e^{-\frac \nu 2-\lambda
}\left( 1+q\right) ,  \label{pressu}
\end{equation}
where $q\equiv p/\rho _0$, $\stackrel{\sim }{\sigma }\equiv \sqrt{8\pi /\rho
_0}\sigma $, $\stackrel{\sim }{\chi }\equiv \sqrt{8\pi /\rho _0}\chi $, and
dot denotes differentiation with respect to $\stackrel{\sim }{x}$. For
numerical calculation, starting from the centre of the configuration $%
\stackrel{\sim }{x}=0$ we should specify values of $\stackrel{\sim }{\sigma }%
\left( 0\right) $, $\stackrel{\sim }{\chi }\left( 0\right) $ and $q\left(
0\right) $ besides $\nu \left( 0\right) $ and $\lambda \left( 0\right) $
(see eq. (\ref{boundcond0})). Obviously, $\stackrel{\sim }{\sigma }\left(
0\right) =0$. From the equation (\ref{eq2}), using (\ref{masa}), (\ref{dmasa}%
) and (\ref{coord}) we find:

\begin{equation}
\stackrel{\sim }{\chi }=\pm 4\sqrt{q_0}.  \label{boundcond}
\end{equation}

Constants of integration in terms of boundary values of $\nu _s$, $\lambda
_s $, $\stackrel{\sim }{\sigma }_s$and $\stackrel{\sim }{x}_s$, have the
following form:

\begin{equation}
A=e^{\frac{\nu _s}2+\lambda _s}C^{-\frac 12};\;B=\frac{6\sqrt{2\pi \rho _0}%
\stackrel{\sim }{\sigma }_s}{4e^{\frac{\nu _s}2+\lambda _s}+3\stackrel{\sim
}{\sigma }_s\stackrel{\sim }{x}_s};\;C=\left( e^{\frac 34\lambda _s}+\frac 34%
\stackrel{\sim }{\sigma }_s\stackrel{\sim }{x}_se^{-\frac{\nu _s}2-\frac{%
\lambda _s}4}\right) ^{\frac 34},  \label{ABC}
\end{equation}

\begin{equation}
a=e^{\frac{\nu _s}2}-\frac 14\stackrel{\sim }{\sigma }_s\stackrel{\sim }{x}%
_se^{-\lambda _s};\;b=\sqrt{\frac{\pi \rho _0}2}\stackrel{\sim }{\sigma }%
_se^{-\lambda _s};\;c=e^{\lambda _s}.  \label{abc}
\end{equation}

The numerical calculation of the equations (\ref{equt1}) - (\ref{pressu})
was carried out for configurations with the parameter values $10^{-2}\leq
q_0\leq 1$. The results of numerical calculation are given on figures.

\begin{figure}[tbph]
\begin{center}
\epsfig{figure=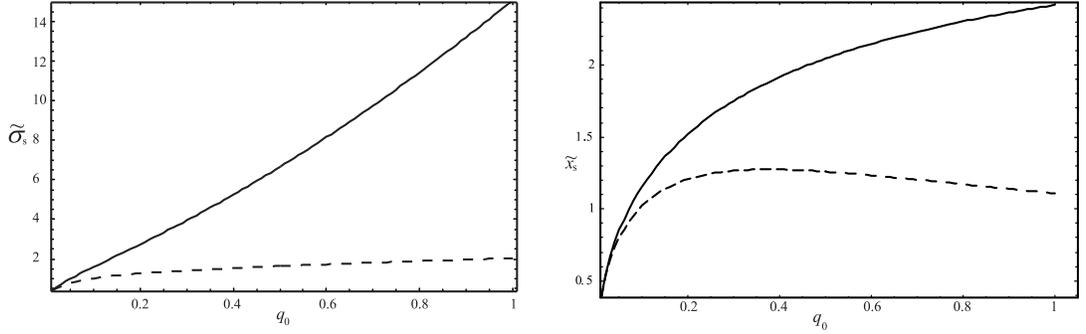,width=14.32cm,height=4.44cm}
\end{center}
\caption{The dependence of "surface density" (on the left) and of "proper
thickness" of the plate (on the right) on the parameter $q_0$ (dotted curve
corresponds to the case of $\stackrel{\sim }{\chi }\left( 0\right) <0$, and
continuous curve -$\stackrel{\sim }{\chi }\left( 0\right) >0$ ).}
\label{fig1}
\end{figure}
\begin{figure}[tbph]
\begin{center}
\epsfig{figure=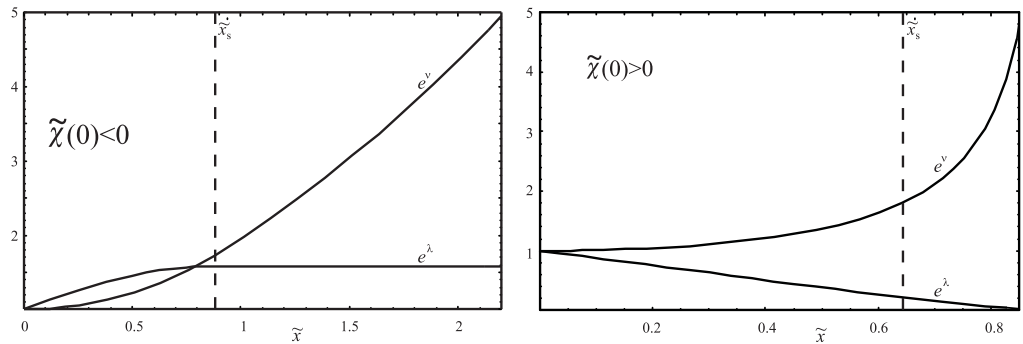,width=13.49cm,height=4.52cm}
\end{center}
\caption{Dependence of functions $e^{\nu (\stackrel{\sim }{x})}$ and $%
e^{\lambda (\stackrel{\sim }{x})}$ on the dimensionless coordinate $%
\stackrel{\sim }{x}$. The dotted vertical line is the bound of the
configuration.}
\label{fig2}
\end{figure}

{\large External solutions.} We study now external solutions (\ref{Taybsol})
and (\ref{oursol}).

In Taub solution (\ref{Taybsol}) constant $B>0$ (as $\lambda ^{\prime
}(x_s)<0$), therefore, on final distance from the centre of the plate in the
point $x=1/B$ the metric has real singularity: $\left| g_{ik}\right| =0$.
And, it can be easily convinced that $x_s<1/B$, i.e. the plate is inside of
the singular planes. Indeed, from the equations it can be seen that the
function $\lambda (x)$ monotonically decreases in all space ($\lambda
^{\prime }<0$ and $\lambda ^{\prime \prime }<0$) and becomes ''$-\infty $ ''
only outside the plate, as function is a finite quantity inside the plate.

In the solution (\ref{oursol}) constant of integration $b$ is also positive,
as $\nu ^{\prime }(x_s)>0$. The solution is 4-flat, what can be convinced by
evaluating Riemannian tensor for this metric. However, it can be checked
easier using transformations

\begin{equation}
a+bx=b\sqrt{x_1^2-t_1^2};\;bt=arcth\frac{t_1}{x_1}.  \label{trans}
\end{equation}
Thus, the metric (\ref{oursol}) takes the Galilean form

\begin{equation}
ds^2=dt_1^2-dx_1^2-dy^2-dz^2,  \label{galilean}
\end{equation}

Thus, this solution represents realization of a homogeneous gravitational
field in sense of GR, which can be excluded by one global transformation (%
\ref{trans}) in all space outside the plate.

{\large Conclusions.} It is conventionally assumed that in the Einstein's
theory the presence of gravitational field in some area is indicated by
nonzero Riemannian tensor in this area. However, the analysis of this
solution leads to the following conclusion: {\it the metric (\ref{oursol})
describes a homogeneous gravitational field of the plate in reference
system, fixed with respect to the plate, but for it the Riemannian tensor }$%
R_{klm}^i${\it \ is zero.} Really, in this field the trial particle moves
under the law

\begin{equation}
a+bx=\frac{a+bx_0}{\cosh bt}  \label{geodez}
\end{equation}
and falls on the plate in finite time $t_0=\arccos {\rm h\,}(a+bx_0/a+bx_s)$%
, where $x_0>x_s$ is the coordinate of the point, where the particle was at
rest at the moment $t=0$.

We think that namely this solution (but not Taub solution) is a relativistic
analogue of the appropriate solution in the Newton's theory.

\end{document}